\documentclass[prl,superscriptaddress,showpacs,twocolumn]{revtex4}

\usepackage{graphicx}

\usepackage{graphicx,amsfonts}

\usepackage{epsfig,amsmath}

\usepackage{verbatim}

\begin{document}

\newcommand{\atanh}
{\operatorname{atanh}}

\newcommand{\ArcTan}
{\operatorname{ArcTan}}

\newcommand{\ArcCoth}
{\operatorname{ArcCoth}}

\newcommand{\Erf}
{\operatorname{Erf}}

\newcommand{\Erfi}
{\operatorname{Erfi}}

\newcommand{\Ei}
{\operatorname{Ei}}


\title{Universal non stationary dynamics at the depinning transition}

\author{Alejandro B. Kolton}
\affiliation{CONICET, Centro At\'omico Bariloche, 8400 S. C. de Bariloche, Argentina}
\author{Gr{\'e}gory Schehr}
\affiliation{Laboratoire de Physique Th\'eorique (UMR du
CNRS 8627), Universit\'e de Paris-Sud, 91405 Orsay Cedex, France}
\author{Pierre Le Doussal}
\affiliation{CNRS-Laboratoire de Physique
Th\'eorique de l'Ecole Normale Sup\'erieure, 24 Rue Lhomond 75231
Paris, France}

\date{\today}

\begin{abstract}

We study the non-stationary dynamics of an elastic
interface in a disordered medium  
at the depinning transition. We compute the two-time response and
correlation functions, found to be universal and
characterized by two independent critical exponents. We find a good agreement 
between two-loop Functional Renormalization Group calculations and
molecular dynamics simulations for the scaling forms, and
for the response aging exponent $\theta_R$. We also describe a dynamical
dimensional crossover, observed at long times  
in the relaxation of a finite system. Our results are 
relevant for the non-steady driven dynamics 
of domain walls in ferromagnetic films and contact lines in wetting.
\end{abstract}


\maketitle

The universal glassy properties that emerge from the 
frustrating competition between elasticity and disorder
are relevant for many experimental
systems, such as interfaces describing  
magnetic~\cite{lemerle_domainwall_creep,
repain_avalanches_magnetic,metaxas_magneticwall,lee4}
and ferroelectric~\cite{paruch_ferro_roughness_dipolar,tybell_ferro_creep}
domain walls, contact lines of fluids~\cite{moulinet_distribution_width_contact_line2,contactfrg}
and fracture~\cite{ponson_fracture,alava_fracture}.
Disorder leads to pinning, affecting in a dramatic way their dynamical
properties. In particular, when driven by an external 
force $f$ at zero temperature, disorder leads to a depinning transition at 
a threshold value $f=f_c$, below which the interface is immobile, and 
above which steady-state motion sets in. For $f \gtrsim f_c$ it 
has been fruitful to 
regard the depinning transition as a critical phenomenon, with 
the mean velocity $v$ as an order 
parameter, $v \sim (f-f_c)^\beta$, and with a  
characteristic length $\xi$ playing the role of the divergent 
correlation length $\xi \sim (f-f_c)^{-\nu}$, 
$\beta$ and $\nu$ being universal
exponents~\cite{fisher_depinning_meanfield}.
Near the critical point, however, the time needed to reach such 
a~\emph{non-equilibrium} steady-state can be very long, since 
the memory of the initial condition persists for length scales larger
than a growing  
correlation length $\ell(t) \sim t^{1/z}$, with $z$ the dynamical
exponent~\cite{gregpld_epl,kolton_line_short_time}.  
Being only limited by the divergent steady correlation 
length $\xi$ or the system size $L$, the resulting~\emph{non-steady} 
critical regime is macroscopically large, $t \lesssim \xi^z,
L^z$. It is thus relevant for experimental protocols.
Analogously to non-driven systems relaxing 
to their critical equilibrium states~\cite{leto_leshouches,
  calabrese_review}, we show here that the transient dynamics
of a \emph{driven} disordered system displays interesting, though different, 
universal features.
\begin {figure}
\includegraphics[width=8cm]{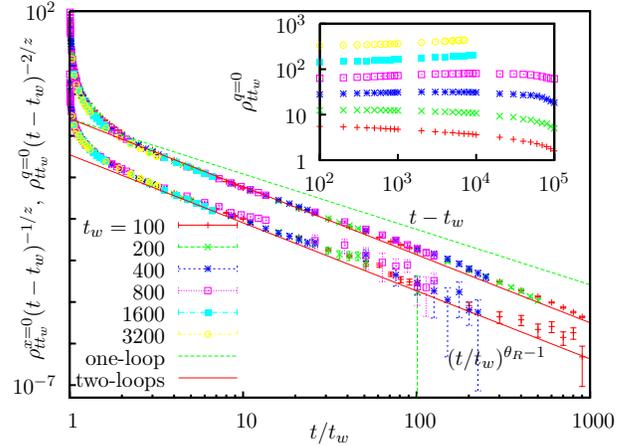}
\caption{Scaling of the local $\rho^{x=0}_{tt_w}$ and the center of
  mass $\rho^{q=0}_{tt_w}$ integrated linear response functions of an
  elastic string relaxing at its depinning threshold (they have been shifted for clarity). 
The new critical
  exponent $\theta_R = - 0.6 \pm 0.05$ numerically obtained is in good
  agreement with our two-loop FRG prediction $\theta_R \approx
  - 0.64$~(\ref{theta_twoloops}). Here $z=1.5$ is the dynamical exponent,
  estimated numerically. Inset: non-scaled 
  response data for the center of mass.} 
\label{fig1}
\end {figure} 

Dynamical properties are characterized by two-time $t,t_w$ 
response and correlation functions, describing the time evolution of the
system as a function of its ``age'' or waiting-time $t_w$ from a given initial condition at
$t=0$. Progress was achieved in the understanding of 
the steady-state, where these functions depend only on 
$t-t_w$. Functional Renormalization Group (FRG) 
calculations~\cite{natterman,narayan_fisher_depinning,chauve_creep_long,ledoussal_frg_twoloops} 
allowed to compute the critical exponents describing 
different universality classes, and powerful algorithms were developed
to elucidate the low temperature dynamical phase diagram 
~\cite{kolton_depinning_zerot2}. In contrast, little is known about the more
difficult transient regime,  
where the time-translation invariance is broken. Yet the first steps in 
that direction have unveiled rich and universal behaviors, including slow 
dynamics~\cite{kolton_line_short_time} and  
aging properties characterized by new exponents~\cite{gregpld_epl}.

Here we focus on elastic interfaces
of dimension $d$ ($d=1$ for an elastic line) 
parameterized by a scalar field $u_{x,t}$ 
describing their position in a $d+1$-dimensional disordered medium. The 
driven overdamped dynamics of this model system
obeys the equation of motion  
\begin{equation} \label{eq_dyn}
\eta \partial_ t u_{x,t} = c \nabla^2 u_{x,t} + F(x, u_{x,t})+ f \;,  
\end{equation}
where $\eta$ is the friction coefficient, $c$ the elastic constant, and 
$F(x,u)$ a quenched random pinning force with disordered averaged
correlations $\overline{F(x, u) F(x', u')}= \Delta(u-u')
\delta^d(x-x')$. Under an applied force $f$, the 
velocity is $v = L^{-d} \int d^d x \overline{\partial_t u_{x,t}}$. In 
this paper we consider a flat initial configuration $u_{x,t=0}=0$ but our 
results hold for any short ranged correlated initial conditions.  
Denoting $\hat u_{q,t}$ the spatial Fourier transform of $u_{x,t}$, 
we focus on the linear response 
${\cal R}^q_{tt_w}$ to a small external
field ${\hat h}_{-qt_w}$ and the correlation function ${\cal C}^q_{tt_w}$:
\begin{equation}\label{def_R}
{\cal R}^q_{tt_w} = \overline{ \delta
  {\hat u}_{qt}/\delta 
  {\hat h}_{-qt_w}}  \quad, \quad {\cal{C}}^q_{tt_w} = \overline{{\hat u}_{qt}
      {\hat u}_{-qt_w}} \;. 
\end{equation}

The main result of this Letter is to establish, both via numerical calculations and
additional analytical work, that these central observables take the scaling form:
\begin{eqnarray}
\label{scaling_resp} {\cal R}^q_{tt_w} &=& \left(t/t_w\right)^{\theta_R} q^{z-2} F_R[q^z
(t-t_w),t/t_w] \;,   \\
\label{scaling_corr}
{\cal C}^q_{tt_w} &=& q^{-(1+2\zeta)} (t/t_w)^{\theta_C-1} F_C[q^z
(t-t_w), t/t_w] \;,
\end{eqnarray}
with $\theta_R$ and $\theta_C$ two new universal 
critical exponents, i.e. independent of
the usual depinning exponents. These are defined such that 
$F_{R,C}(y_1,y_2\to \infty) \sim f_{R,C}(y_1)$ for fixed $y_1$. In
Eq. (\ref{scaling_resp}), (\ref{scaling_corr}), $F_{R,C}$ are
universal scaling functions (up to a non universal amplitude) and
$\zeta$ the roughness exponent. In the
limit $y_1\to 0$, $y_2$ fixed,  
one finds $F_R(y_1,y_2)\sim y_1^{(2-z)/z}g_R(y_2)$, $F_C(y_1,y_2)\sim
y_1^{(d+2\zeta)/z}g_C(y_2)$, {\it i.e.} a well defined $q\to 0$ limit. 
These scaling forms were predicted in Ref.~\cite{gregpld_epl} 
based on a one-loop FRG calculation. One may question however
whether this lowest order in the $d=4-\epsilon$ dimensional expansion 
is accurate enough to describe interfaces of experimental
interest $d=1,2$. In addition, no prediction for $\theta_C$ was obtained.
Here we firmly establish that the above scaling forms hold and we
provide a reliable determination of $\theta_R$ and $\theta_C$ in $d=1$. 
We also perform a two-loop FRG calculation, as is known to be
required for a consistent theory of depinning~\cite{ledoussal_frg_twoloops}.
   
Most of the numerical studies of the transient dynamics have focused 
so far on one-time quantities which can be obtained from ${\cal{C}}^q_{tt'}$
and ${\cal{R}}^q_{tt'}$ in Eq. (\ref{scaling_resp}) and (\ref{scaling_corr}). 
The structure factor $S_q(t) \equiv {\cal{C}}^q_{tt}$ was found ~\cite{kolton_line_short_time} 
to behave as:\begin{equation}
S_q(t) \equiv {\cal{C}}^q_{tt} 
\sim q^{-(d+2 \zeta)} F[q {\ell}(t)] \;, \; \ell(t) \sim t^{1/z} \;,
\label{strucfac}
\end{equation}
where $F(y)\sim c^{\rm st}$, a constant, for $y \gg 1$ and $F(y)\sim
y^{d+2\zeta}$ for $y \ll 1$. The  relaxational dynamics is thus 
dictated by a single growing 
length, separating the small, steady-state equilibrated scales, 
from the large ones retaining a long-time memory of the initial 
condition. Eq. (\ref{strucfac}) is obtained from (\ref{scaling_corr}) in the limit $t \to t_w$ ({\it i.e.}
$y_2\to 1$, $y_1 \to 0$)  
with $q^z t$ ({\it i.e.} $y_1 y_2/(y_2-1)$) fixed. The
analogy with standard critical phenomena suggests
for the velocity, the scaling form, 
$v(t,f) = b^{-\beta/\nu} G[b^{-z}t, b^{1/\nu} (f-f_c), b^{-1} L]$ 
where $b$ is an arbitrary rescaling factor, numerically verified in Ref.~\cite{kolton_line_short_time}. 
For $f=f_c$ and $t \ll L^z$ it implies that 
$v(t) \propto t^{-\beta/\nu z}$ and also that:
\begin{eqnarray}
dv(t)/df \propto A \; t^{(2-z)/z} \;,
\label{dvdfvst}
\end{eqnarray}
where we used the exact relations $\beta = \nu(z-\zeta)$ 
and $\nu = 1/(2-\zeta)$, from statistical tilt symmetry (STS)
~\cite{narayan_fisher_depinning}. We now check that this
scaling of one time observables for $f=f_c$  
is consistent with the two time scaling Eq.~(\ref{scaling_corr}).
Indeed Eq.~(\ref{dvdfvst}) results by combining 
the  {\it exact} relation~\cite{chauve_creep_long}
$\frac{d}{df} v(t) = \int_0^t ds\; \partial_t {\cal R}^{q=0}_{ts}$ 
and the limit $q\to 0$ of Eq.~(\ref{scaling_resp}) 
\begin{eqnarray}
{\cal R}^{q=0}_{tt_w} \sim (t-t_w)^{(2-z)/z} \left(t/t_w\right)^{\theta_R} g_R(t/t_w) \;,\label{scaling_resp_zeromode}
\end{eqnarray}
with $g_R(x) \propto c^{\rm st}$ a constant for $x \gg 1$. 

Let us now focus on two time quantities. To check
Eqs.~(\ref{scaling_resp}) and (\ref{scaling_corr})  
we have performed numerical simulations of Eq.~(\ref{eq_dyn}) in the
case of elastic lines, $d=1$, experimentally relevant for many two
dimensional systems, e.g. films. 
To study the non-stationary dynamics at $f=f_c$ we discretize
Eq.~(\ref{eq_dyn}) in the $x$ direction, $u_{x,t}\equiv u_i(t)$, with
$i=0,...,L-1$, and use the method described in
Ref.~\cite{kolton_line_short_time}. We start at $t=0$ with a flat
configuration, $u_i(t=0)=0$, and monitor correlation and response
functions at the exact sample critical force
$f_c$
~\cite{rosso}. Numerically, it is more convenient to work
with the local integrated response  
$\rho^{x=0}_{tt_w} \equiv \int_0^{t_w} ds\;\int_q {\cal R}_{ts}^q$
(where $\int_q$ denotes the integral over the first Brillouin zone)
and zero-mode integrated response $\rho^{q=0}_{tt_w} \equiv
\int_0^{t_w} ds{\cal R}_{ts}^{q=0}$. From Eq.~(\ref{scaling_resp}) we
predict, 
\begin{equation}
\frac{\rho^{x=0}_{tt_w}}{(t-t_w)^{1/z}} = h\left(\frac{t}{t_w}\right), \; 
\frac{\rho^{q=0}_{tt_w}}{(t-t_w)^{2/z}} = \tilde{h}\left(\frac{t}{t_w}\right)
\;, 
\label{predicted_responses}
\end{equation}
where both $h(y)$ and $\tilde h(y)$ behave as $y^{-1+\theta_R}$ for $y\to \infty$.  
To implement the local (zero mode) response we define the observable 
$w_i(t) = u_i(t) - u_{cm}(t)$ ($w_i(t) = u_i(t)$) where $u_{cm}(t)\equiv \frac{1}{L} \sum_i u_i(t)$ and then compute,
\begin{equation} 
\rho^{x=0 (q=0)}_{tt_w} = \lim_{\alpha \rightarrow 0} \frac{1}{L} 
\sum_i \overline{[w_i(t) - w_i^{\alpha}(t)]\sigma_i} \alpha^{-1} , 
\end{equation}
where $w_i^{\alpha}(t)$ is the solution of Eq.~(\ref{eq_dyn}) with
$u_i^{\alpha}(0)=u_i(0)=0$ and an additional perturbative force $\alpha
\sigma_i \theta(t_w-t)$. We take random numbers $\sigma_i = \pm 1$  
uncorrelated from site to site for computing $\rho^{x=0}_{tt_w}$, 
and $\sigma_i = \sigma_0$ for computing $\rho^{q=0}_{tt_w}$.
The value of $\alpha$ is chosen small enough to guarantee linear response~\cite{greg_superrough}.
In Fig.~\ref{fig1} we show the numerical results for $\rho^{x=0}_{tt_w}$ and
$\rho^{q=0}_{tt_w}$, for $L=2048$ averaged over 10000 disorder
realizations. We see that the predicted scaling forms,
Eq.~(\ref{predicted_responses}), describe well the data.  
For $t/t_w \gg 1$ we observe 
a well developed power law behavior with an aging exponent $\theta_R=-0.6 \pm
0.05$ 
which is indistinguishable for both responses,
$\rho^{x=0}_{tt_w}(t-t_w)^{-1/z}\sim \rho^{q=0}_{tt_w}(t-t_w)^{-2/z} 
\sim (t/t_w)^{-1+\theta_R}$, as
predicted. 
How does this numerical estimate for
$\theta_R$ compare with the previous FRG approach of Ref.~\cite{gregpld_epl}?
The one loop result 
for $\theta_R =  -\frac{\epsilon}{9} + {\cal O}(\epsilon^2)$, 
setting $\epsilon = 3$ gives $\theta_R \approx
-1/3$. Incidentally, up to one loop order, one finds the relation
$\theta_R = (z-2)/z$, which, using the numerical estimate $z =
1.5$~\cite{kolton_line_short_time}, yields again 
$\theta_R \approx -1/3$. Although it goes in the
right direction, it is still far from our numerical result. To see
whether the FRG predictions can be improved we have 
computed ${\cal R}^{q=0}_{tt_w}$ up to two loop order~\cite{tbp}. 
The starting point is Eq. (\ref{eq_dyn}). Response and correlations are then
obtained from the standard dynamical (disorder averaged)
Martin-Siggia-Rose action ${\cal S}$ which reads here  
\begin{eqnarray}\label{def_action}
&&{\cal S} = \int_{t>t'>0}  \int_q i \hat u_{qt} [(q^2 +  \partial_t ) \delta_{tt'} + \hat \Sigma_{tt'}] u_{-qt'} \nonumber \\
&&- \frac{1}{2} \int_{x,t>t'>0} i \hat u_{xt} i \hat u_{xt'} \Delta(u_{xt} -
u_{xt'}) \;,
\end{eqnarray}
where $\Delta(u)$ is the force-force correlator and $\hat
\Sigma_{tt'}$ is the self-energy. As
a result of the covariance of the action under STS \cite{narayan_fisher_depinning} the self-energy has
the structure  
$\hat \Sigma_{tt'} = \Sigma_{tt'} - \delta_{tt'} \int_0^t dt_1
\Sigma_{tt_1}$. It was computed to one loop in
Ref. \cite{gregpld_epl} and at two loop order the
perturbation theory leads to diagrams similar to the one 
contributing the dynamical exponent $z$, as
depicted in Fig. 10 of Ref. \cite{ledoussal_frg_twoloops}, with the
constraint that here, the time variables are positive.   
The (bare) response function ${\cal R}^q_{tt'} = \langle
i\hat u_{qt} u_{-qt'} \rangle_{\cal S}$ is then computed from the
exact identity
\begin{eqnarray}
&&{\cal R}^q_{tt'} = R^q_{tt'} - \int_{t'<t_1<t_2<t} R^q_{tt_2}
  \Sigma^q_{t_2 t_1} 
{\cal R}^q_{t_1t'} \nonumber \\
&&+ \int_{t'<t_1<t} R^q_{tt_1} {\cal R}^q_{t_1t'}
[\int_{0<t_2<t_1} \Sigma^q_{t_1 t_2}] \;,
\end{eqnarray}
where $R^q_{tt'} = \theta(t-t')e^{-q^2(t-t')}$ is the response in the absence
of disorder.   
\begin{figure}[h]
\begin{center}
 \includegraphics[width=8cm]{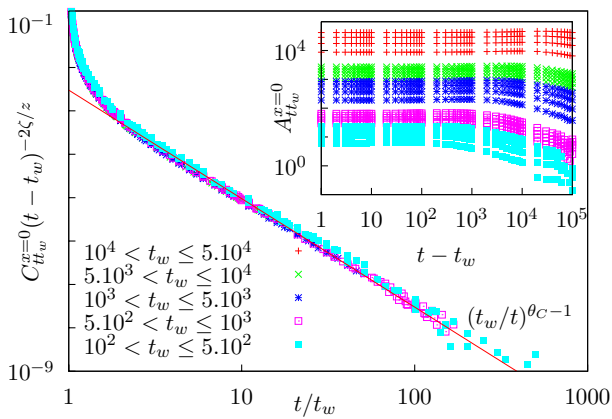}
\caption{Scaling of the local autocorrelation function. The  
aging exponent is $\theta_C=-1.5 \pm 0.05$, different from $\theta_R$. 
Inset: non-rescaled data.}\label{fig2}
\end{center}
\end{figure}
Reexpressing in terms of $\Delta(u)$, corrected to the same order, and
using the FRG fixed point equation, one explicitly shows that it
has the scaling form as in Eq. (\ref{scaling_resp_zeromode}).
We then find that no new independent divergence occurs in $t/t'$ at this order, hence that
to two loop accuracy the relation $\theta_R = (z-2)/z + {\cal O}(\epsilon^3)$
continues to hold. Our numerical result however indicates that this relation cannot 
hold to all orders in $\epsilon$, i.e one must have $\theta_R \neq (z-2)/z$, implying that $\theta_R$ is
indeed a new independent exponent. One way to understand the
FRG result is then to rewrite more explicitly, 
\begin{eqnarray}\label{theta_twoloops}
\theta_R  &=&  -\frac{\epsilon}{9} + \left(\frac{1}{162 \gamma \sqrt{2}} -
\frac{\log{2}}{108} - \frac{23}{648} \right)\epsilon^2 +
O(\epsilon^3) \nonumber 
\\
&=& -0.1111... \epsilon - 0.03395... \epsilon^2 + O(\epsilon^3) \;,
\end{eqnarray}
with $\gamma = 0.54822..$. 
We note that if we set $\epsilon=3$ in that expression (\ref{theta_twoloops})
assuming the $O(\epsilon^3)$ to be small, we obtain $ \theta_R = -0.64...$, very close
to the the numerical value. Hence we conclude that although corrections
to 3 loop and higher to $\theta_R-(z-2)/z$ must be large, in $\theta_R$ they
must be small. This provides one way of interpreting our results, and
motivation for future analytical work.

We have also checked numerically the scaling form for the correlation
function in Eq. (\ref{scaling_corr}). It is 
more convenient to compute the autocorrelation function ${\cal
  C}^{x=0}_{tt_w} = \overline{u_{x,t} u_{x,t_w}}$
obtained from 
${\cal C}^{q}_{tt_w}$ by integration over Fourier modes ${\cal
  C}^{x=0}_{tt_w}  = \int_q {\cal C}^{q}_{tt_w}$. From 
Eq. (\ref{scaling_corr}), one expects ${\cal C}^{x=0}_{tt_w}  \sim  (t-t_w)^{2 \zeta/z} \hat
h\left(t/t_w\right)$
where $\hat h(y) \propto y^{-1+\theta_C}$ for large $y$. To check this,
we compute numerically ${\cal
  C}^{x=0}_{tt_w} = L^{-1} \sum_i \overline {w_i(t)w_i(t')} $.  
In Fig.~\ref{fig2} we show a plot of $(t-t_w)^{-2\zeta/z}{\cal
  C}^{x=0}_{tt_w}$ for $L=2048$ obtained by averaging over 10000 samples,
which is in good agreement with the predicted scaling. For $t/t_w \gg 1$,   
we see a power-law behavior with a non-equilibrium exponent
 $\theta_C = -1.5 \pm 0.05$. At variance with pure critical
 dynamics~\cite{calabrese_review}, one obtains that $\theta_C$
 and $\theta_R$ are~\emph{different} exponents. Such
behavior was observed in other disordered elastic systems 
(though relaxing at equilibrium) 
 \cite{co_pre}. The determination of $\theta_C$ via the FRG
requires a computation to order ${\cal O}(\epsilon^2)$ of ${\cal C}^q_{tt_w}$ 
and remains a challenge.

So far we have analyzed the situation where $\ell(t) < L$. 
Finite-size effects can be observed in experiments however, 
as shown recently for domain walls in magnetic nanowires~\cite{lee4}, and 
vortex lattices in micron-sized superconductors~\cite{moira}. 
The finite-size crossover manifests in the velocity as
$v(t)L^{\beta/\nu}=f(t/L^z)$ and for $t>L^z$ the interface losts memory of the initial 
condition~\cite{kolton_line_short_time}. 
\begin{figure}[h]
\includegraphics[width=\linewidth]{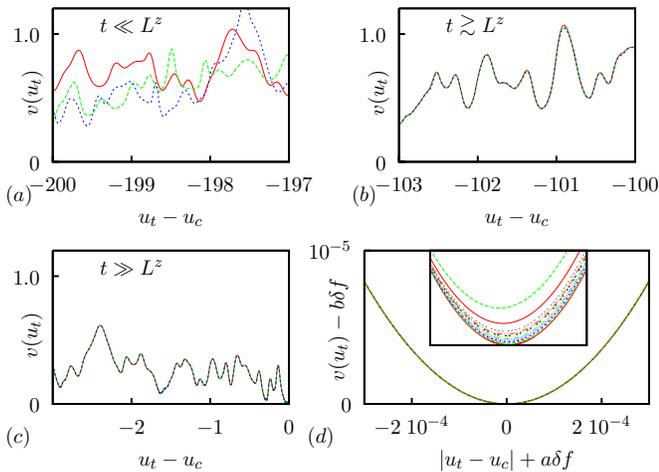}
\caption{(a-c) Three different initial conditions evolving at $f_c$ in the
  same sample coalesce into a unique reparametrized velocity function $v(u_{cm}(t))$ after a typical time $t\sim L^z$, before 
stoping at the 
critical configuration, $u_c \equiv u_{cm}(\infty)$. (d) 
The steady-state structure 
of $v(u_{cm})$ around $u_c$ for different forces $\delta f \equiv f-f_c \to
  0^+$  
is a well defined parabola with a positive (negative) shift proportional to $\delta f$ in the 
position (velocity) axis. Inset: unshifted data. One finds for $L=32$
  and $\delta f/f_c = 6-13,20,30\;10^{-6}$ 
$a \approx 0.16$, $b \approx 1.45$. 
}
\label{fig3}
\end{figure}
What happens next? Remarkably, a detailed analysis of the dynamics 
for $t>L^z$ reveals that different initial conditions in the same
sample evolve collapsing into a unique, sample-dependent, reparametrized velocity $v(u_{cm})$ (as a function of
the center of mass position),
before stopping at the critical configuration, as shown in 
Fig.~\ref{fig3}(a-c). This behaviour can be seen as a direct consequence of the Middleton 
theorems, which assures the convergence to a unique periodic attractor 
for $f \geq f_c$~\cite{middleton_theorem}. 
For $t \gg L^z$ we can thus describe the velocity by an 
effective equation of motion for a 
particle, $\dot u_{t} = v(u_t,f)$ with 
$u_{t}\equiv u_{cm}(t)$. 

Before exploring further the consequences, let us note that
the $d=0$ problem of a particle is a solvable limit, interesting
per se, as the (sample dependent) response function is given exactly by ~\cite{chauve_thesis} 
$R_{tt_w} = \theta(t-t_w) v(u_t,f)/v(u_{t_w},f)$. In the usual model
$\eta v(u_t,f)=F(u_t)+f$, near the threshold,
$\delta f = f-f_c \ll f_c$, the particle spends most time near the zero
force point (set to be at $u=0$). For a smooth force field we can write 
$\eta \dot u_{t} = \gamma u^2_t + \delta f$ which yields $v(t)\sim t^{-2}$ and 
\begin{eqnarray}
R_{tt_w} = \theta(t-t_w) 
\begin{cases}
\frac{\sin^2{(t_w \sqrt{\gamma \delta f}/\eta)}}{\sin^2{(t \sqrt{\gamma \delta f}/\eta)}}
\; , \; \delta f > 0 \\
\frac{\sinh^2{(t_w \sqrt{\gamma |\delta f|}/\eta)}}{\sinh^2{(t \sqrt{\gamma |\delta f|}/\eta)}}
\; , \; \delta f < 0 \\
\left(t/t_w\right)^{-2}
\end{cases}
\end{eqnarray}
for $\delta f \to 0$. Hence $\theta_R(d=0) = -2$, consistent with a monotonic dependence of
$\theta_R$ with $d$, and $\beta(d=0)=1/2$.

Next, we confirm that the results of the $d=0$ model are relevant for the interface for $\ell(t)>L$.
As shown in Fig. \ref{fig3}d, we checked that for 
interfaces of sizes $L=32$ and small $\delta f \ll f_c$ the reparametrized velocity has a nice
parabolic shape near the zero force point, $v(u,f) = \gamma (u + a \delta f)^2 + b \delta f$
where the constants $b,\gamma>0$ (their size dependence will be studied
elsewhere \cite{tbp}). $a$ being found irrelevant, this result is 
consistent with the steady-state value $\beta=1/2$ found for the interface in the regime $\delta f^{-\nu} \gg L$~\cite{duemmer_crossover},
and predicts a crossover to an effective $\theta_R^{eff}= -2$ in the fixed $L$, large $t$ limit
 for the interface. 

For modeling contact lines ~\cite{moulinet_distribution_width_contact_line2} the term $\nabla^2 u_{x,t}$ in Eq.(\ref{eq_dyn}) is
replaced by a long range elastic force $\int_{x'} [u(x',t)-u(x,t)]/|x-x'|^2$. 
In this case, using the same numerical method, we confirm the scaling forms (\ref{scaling_resp})
(replacing $q^{z-2} \to q^{z-1}$) and measure  
$\theta_C= -1.2 \pm 0.1$ and  $\theta_R= -0.5 \pm 0.1$. The
same analysis leading to (\ref{theta_twoloops}) gives $\theta_R=-0.22$ to one
loop and $\theta_R=-0.38$ to two loop. 

To conclude we have confirmed numerically the scaling forms for non stationary dynamics at depinning,
for model systems of experimental relevance. The exponent $\theta_R$ is found in reasonable
agreement with FRG predictions. An interesting dimensional crossover was found at large
time. We hope this motivates new experiments, e.g in magnets and wetting.
 
This work was supported by the France-Argentina MINCYT-ECOS A08E03. 
A.B.K aknowledges the hospitality at LPT-Orsay and ENS-Paris.

\end{document}